\documentclass[twocolumn,aps,prd,amsmath,amssymb]{revtex4-2}

\usepackage{graphicx}
\usepackage{xcolor}
\usepackage{dcolumn}
\usepackage{bm}
\usepackage{amsthm,physics,hyperref,mathtools,dsfont}
\usepackage[capitalise]{cleveref}
\usepackage{bbm}

\begin{document}

\title{Operator Entanglement from Non-Commutative Symmetries}

\author{Michele Arzano}
\email{michele.arzano@na.infn.it}

\author{Goffredo Chirco}
\email{goffredo.chirco@unina.it}
\affiliation{Dipartimento di Fisica ``E. Pancini", Universit\`a di Napoli Federico II, I-80125 Napoli, Italy}
\affiliation{INFN, Sezione di Napoli, Complesso Universitario di Monte S. Angelo,
Via Cintia Edificio 6, 80126 Napoli, Italy}

\date{\today}

\begin{abstract}
We argue that Hopf-algebra deformations of symmetries---as encountered in non-commutative models of quantum spacetime---carry an intrinsic content of \emph{operator entanglement} that is enforced by the coproduct-defined notion of composite generators.
As a minimal and exactly solvable example, we analyze the $U_q(\mathfrak{su}(2))$ quantum group and a two-qubit realization obtained from the coproduct of a $q$-deformed single-spin Hamiltonian.
Although the deformation is invisible on a single qubit, it resurfaces in the two-qubit sector through the non-cocommutative coproduct, yielding a family of intrinsically nonlocal unitaries. 
We compute their operator entanglement 
in closed form and show that, for Haar-uniform product inputs, their entangling power is fully determined by the latter.
This provides a concrete mechanism by which non-commutative symmetries enforce a baseline of entanglement at the algebraic level, with implications for information dynamics in quantum-spacetime settings and quantum information processing.
\end{abstract}

\maketitle

\section{Introduction}

In standard quantum information theory, entanglement is a \emph{resource} that can be generated and manipulated by suitable local and nonlocal operations~\cite{RevModPhys.81.865,RevModPhys.91.025001, PhysRevLett.86.544}. Its characterization is always relative to a given tensor-product structure, and, for fixed bipartitions, local unitary changes of basis do not affect entanglement measures. Only additional physical constraints can obstruct the ability to remove or generate correlations by allowed operations.

Quantum-spacetime scenarios suggest a complementary perspective.
Spacetime noncommutativity is a widely studied framework for modeling putative quantum-gravitational effects by mathematical structures beyond smooth manifolds. A well developed correspondence associates various non-commutative spaces to \emph{deformed symmetries} described by Hopf algebras (quantum groups) \cite{Arzano:2021scz,Madore:1991bw,Majid:1994cy,Chaichian:2004za,Joung:2008mr,Arzano:2014ppa}.
The key new ingredient in these scenarios is a \emph{non-trivial coproduct}, namely a deformation of the Leibniz rule that prescribes how a generator $X$ acts on tensor products of representations:
\begin{equation}
\Delta X = X \otimes \mathbbm{1} + \mathbbm{1} \otimes X .
\label{eq:coproleib}
\end{equation}
When the symmetry algebra is deformed into a quantum group, $\Delta X$ generically deviates from \eqref{eq:coproleib} and becomes non-symmetric under exchange of factors (non-cocommutativity).
In such a setting, the very notion of composite observables is modified: the extension of ``single-system'' generators to multipartite systems is dictated by the non-trivial Hopf algebra structure. As we argue in this work, this can enforce an \emph{operator-level} layer of entanglement, which is not merely a property of states for a fixed tensor product, but is tied to the algebraic rule of composition itself.

The central claim of this work is that \emph{non-cocommutativity of quantum-group symmetries induces an operator entanglement that is enforced by the coproduct-defined notion of composite generators}. We substantiate this claim in the simplest nontrivial setting: the $U_q(\mathfrak{su}(2))$ quantum group and its two-qubit realization in the spin-$\tfrac12$ representation.

\section{Operator entanglement}

Let $\mathcal{H}$ be a $d$-dimensional Hilbert space.
Endowed with the Hilbert--Schmidt inner product
\begin{equation}
\langle X,Y\rangle \coloneqq \Tr(X^\dagger Y),
\end{equation}
the operator space $\mathcal{L}(\mathcal{H})$ becomes a Hilbert space $\mathcal{H}_{HS}$, naturally isomorphic to $\mathcal{H}^{\otimes 2}$.
Given a unitary $U\in\mathcal{L}(\mathcal{H})$, we can associate to it the vector
\begin{equation}
\ket{U} = (U\otimes \mathds{1})\,\ket{\Phi^+},
\quad
\ket{\Phi^+} = \frac{1}{\sqrt{d}}\sum_{i=1}^d \ket{i}\otimes\ket{i},
\label{eq:vecU}
\end{equation}
realizing the Choi--Jamiolkowski isomorphism~\cite{PhysRevA.110.052416}.

For a bipartite structure $\mathcal{H}=\mathcal{H}_A\otimes \mathcal{H}_B$ with $d=d_A d_B$,
the induced vector $\ket{U}$ lives in
$\mathcal{H}^{HS}_{d_A^2}\otimes\mathcal{H}^{HS}_{d_B^2}$ and one defines the \emph{operator entanglement} $E(U)$ as the entanglement of $\ket{U}$ across the $AA'|BB'$ bipartition (where the primed indices refer to the second tensor factor in the definition of $|\Phi^+\rangle$) \cite{PhysRevA.63.040304,PhysRevA.62.030301,PhysRevA.66.044303}.
For linear entropy,
\begin{equation}
E(U)=1-\Tr(\sigma_{AA'}^2),\qquad \sigma_{AA'}=\Tr_{BB'}\big(\ket{U}\bra{U}\big).
\end{equation}
This quantity is invariant under left--right local unitaries of the form
$(U_A\otimes U_B)\,U\,(V_A\otimes V_B)$ and thus probes the intrinsic nonlocality of the \emph{operator} $U$ for a fixed tensor-product structure.

In standard applications the tensor product is a fixed kinematical input.
Here the key additional structure is the coproduct: it dictates how single-system observables extend to two-systems operators.
We will see explicitly that, for $U_q(\mathfrak{su}(2))$, the deformed coproduct injects an irreducible contribution to $E(U)$.

\section{$q$-deformed $\mathfrak{su}(2)$ and Hopf structure}

The $q$-deformed algebra $U_q(\mathfrak{su}(2))$ \cite{Biedenharn:1989jw,Macfarlane:1989dt} is generated by $J_\pm,J_z$ with
\begin{equation}
[J_z,J_\pm]=\pm J_\pm,\qquad [J_+,J_-]=[2J_z]_q,
\end{equation}
where
\begin{equation}
[x]_q=\frac{q^x-q^{-x}}{q-q^{-1}}
\end{equation}
is defined by functional calculus on $J_z$.
The Hopf algebra structure is specified by the coproduct
\begin{align}
\Delta(J_\pm) &= J_\pm\otimes q^{J_z} + q^{-J_z}\otimes J_\pm, \label{eq:coprJpm}\\
\Delta(J_z) &= J_z\otimes \mathds{1} + \mathds{1}\otimes J_z, \label{eq:coprJz}
\end{align}
together with antipodes
\begin{equation}
S(J_z)=-J_z,\qquad S(J_\pm)=-q^{\mp 1}J_\pm .
\end{equation}
In the limit $q\to 1$ one recovers the usual $\mathfrak{su}(2)$ Lie algebra and the cocommutative coproduct.

The crucial point is that for $q\neq1$ the coproduct of $J_\pm$ is \emph{non-cocommutative}:
$\Delta\neq\tau\circ\Delta$ (with $\tau$ the flip map). This non-cocommutativity is the algebraic seed of the Hopf-induced nonlocality explored below.

\section{Single-spin Hamiltonian and the ``accidental'' qubit}
\label{sec:onequbh}

To fix the ideas, let us now consider the following Hamiltonian \cite{Ballesteros:1998yf}
\begin{equation}
H(q)=q^{J_z/2}(J_+ + J_-)q^{J_z/2}.
\label{eq:Hq_def}
\end{equation}
Hermiticity in the standard Hilbert-space inner product requires $q\in\mathbb{R}^+$, which we assume henceforth.

On a general spin-$l$ irrep with basis $\{\ket{l,m}\}$ one finds
\begin{align}
H(q)\ket{l,m}
&= q^{m+1/2}\sqrt{[l-m][l+m+1]}\,\ket{l,m+1}\nonumber\\
&\quad + q^{m-1/2}\sqrt{[l+m][l-m+1]}\,\ket{l,m-1},
\end{align}
so the spectrum and matrix elements depend nontrivially on $q$ for $l>\tfrac12$.
For $q\to1$, $[x]_q\to x$ and $H(1)=2J_x$.

For the fundamental representation $l=\tfrac12$, all $q$-dependence cancels and
\begin{equation}
H(q)=
\begin{pmatrix}
0 & 1\\
1 & 0
\end{pmatrix}
=2J_x
\qquad \forall\,q.
\end{equation}
Thus the deformation is \emph{invisible} at the one-qubit level (see Appendix~\ref{app:single_spin}).
This accidental triviality might suggest that no new quantum-information features arise in this representation; the nontrivial coproduct will show otherwise.

\section{Two-qubit Hamiltonian from the coproduct}
\label{sec:twoqubh}

We now use the coproduct to define a two-qubit Hamiltonian.
In the spin-$\tfrac12$ representation,
\begin{equation}
J_+ = \ket{\uparrow}\!\bra{\downarrow},\quad
J_- = \ket{\downarrow}\!\bra{\uparrow},\quad
2J_z=\sigma_z,
\end{equation}
so
\begin{equation}
q^{2J_z}=\mathrm{diag}(q,q^{-1}),\qquad H(q)=\sigma_x.
\end{equation}
Using that $H(q)$ is linear in $J_\pm$ and applying \eqref{eq:coprJpm}, the coproduct yields (Appendix~\ref{app:hab})
\begin{equation}
H_{AB}(q)= \sigma_x\otimes\mathds{1} + q^{2J_z}\otimes\sigma_x,
\label{eq:HAB_def}
\end{equation}
i.e.
\begin{equation}
H_{AB}(q)=
\begin{pmatrix}
0 & q & 1 & 0\\
q & 0 & 0 & 1\\
1 & 0 & 0 & q^{-1}\\
0 & 1 & q^{-1} & 0
\end{pmatrix}.
\end{equation}
For $q\to1$ one recovers the separable Hamiltonian $H_{AB}(1)=\sigma_x\otimes\mathds{1}+\mathds{1}\otimes\sigma_x$.

The deformation manifests itself entirely at the \emph{two-qubit level}: although each single-spin Hamiltonian is indistinguishable from $\sigma_x$, their composite is genuinely deformed whenever $q\neq1$.

The two-qubit time evolution is determined by the operator
\begin{equation}
U(t)=e^{-itH_{AB}(q)}.
\end{equation}
A key simplifying feature is that $H_{AB}(q)$ satisfies the cubic relation
\begin{equation}
H_{AB}^3(q)=\alpha^2\,H_{AB}(q),
\qquad
\alpha=\frac{q^2+1}{q}=q+q^{-1},
\label{eq:cubic_main}
\end{equation}
with spectrum $\{0,0,\pm \alpha\}$.
Hence any analytic function of $H_{AB}(q)$ reduces to a polynomial of degree at most two.
In particular (Appendix~\ref{app:uni}),
\begin{align}
e^{-itH_{AB}(q)}
&=\mathds{1}-i\frac{\sin(\alpha t)}{\alpha}H_{AB}(q)+\frac{\cos(\alpha t)-1}{\alpha^2}H_{AB}^2(q) \nonumber\\
&=\mathds{1}-i\,\frac{q\,s}{q^2+1}H_{AB}(q)+\frac{q^2(c-1)}{(q^2+1)^2}H_{AB}^2(q),
\label{eq:exp_poly_form}
\end{align}
where $c=\cos(\alpha t)$ and $s=\sin(\alpha t)$.
Explicitly,
\begin{equation}
U(t)
=\frac{1}{q^2+1}
\begin{pmatrix}
1+q^2 c & -i q^2 s & -i q s & q(c-1)\\
- i q^2 s & 1+q^2 c & q(c-1) & -i q s\\
- i q s & q(c-1) & q^2+c & -i s\\
q(c-1) & -i q s & -i s & q^2+c
\end{pmatrix}.
\label{eq:U_matrix}
\end{equation}
This family of unitaries is determined purely by the coproduct deformation: for $q=1$, $U(t)$ factorizes into a product of single-qubit rotations, while for $q\neq1$ it is intrinsically nonlocal.

\section{Operator entanglement}
\label{sec:operent}

For a bipartite unitary $U\in\mathcal L(\mathcal H_A\otimes\mathcal H_B)$ with $\dim\mathcal H_A=d_A$, $\dim\mathcal H_B=d_B$,
the linear-entropy operator entanglement admits the Wang--Zanardi trace form~\cite{PhysRevA.63.040304,PhysRevA.66.044303}
\begin{equation}
E(U)=
1-\frac{1}{d_A^2 d_B^2}\,
\Tr\!\left(
U^{\otimes 2}\,T_{13}\,U^{\dagger\otimes 2}\,T_{13}
\right),
\label{eq:WZ_def}
\end{equation}
where $T_{13}$ swaps the first and third factors in $(AB)\otimes(A'B')\cong A\otimes B\otimes A'\otimes B'$. We use the labelling $(A,B,A',B')\equiv(1,2,3,4)$.

For two qubits ($d_A=d_B=2$),
\begin{equation}
E(U)=1-\frac{1}{16}\Tr\!\left(U^{\otimes 2}T_{13}U^{\dagger\otimes 2}T_{13}\right).
\label{eq:WZ_2qubit}
\end{equation}

\begin{figure}[h]
\centering
\includegraphics[width=1\linewidth]{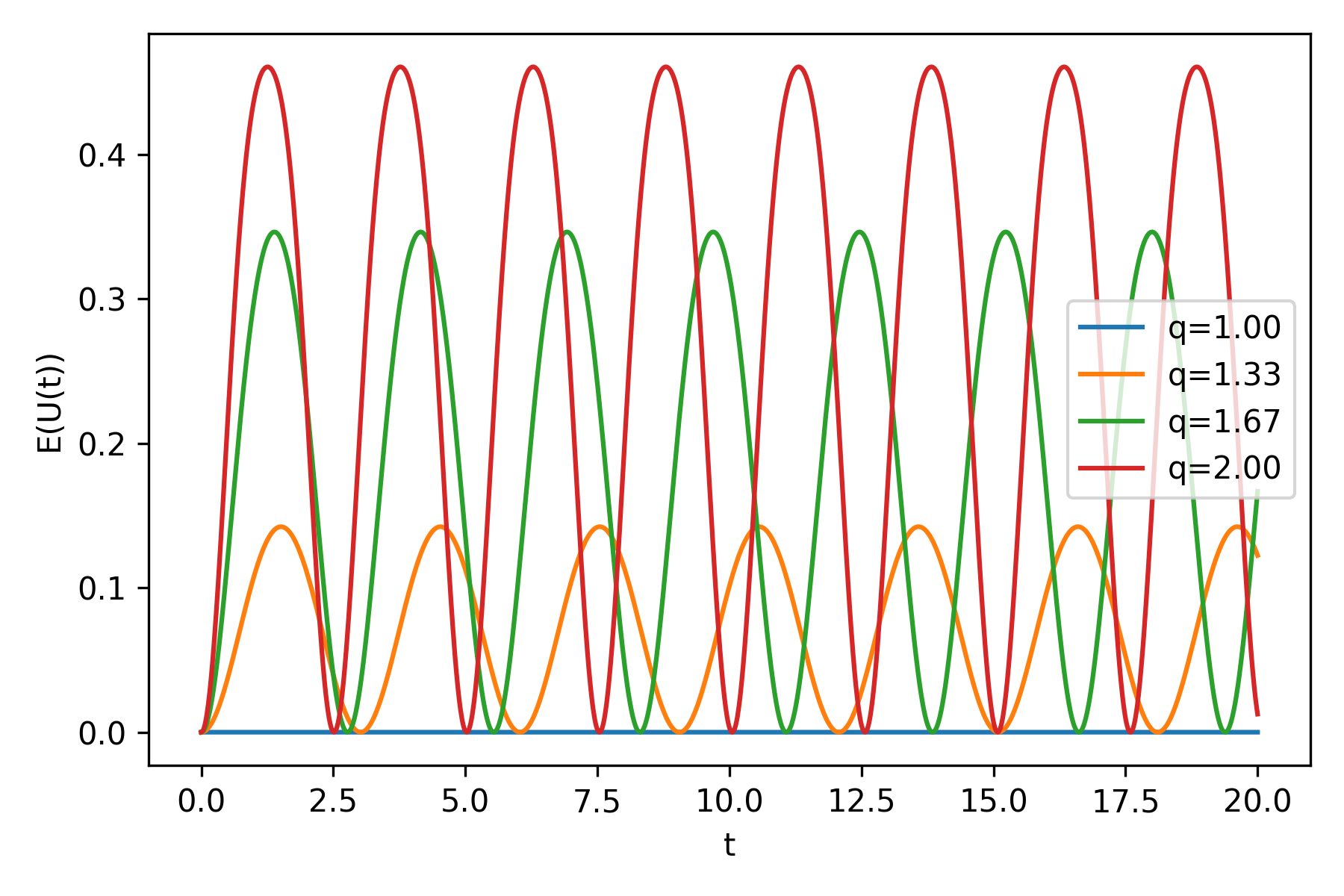}
\caption{Operator entanglement $E(U(t))$ for several values of $q$.
At $q=1$, $E(U(t))=0$.
For $q\neq1$, $E(U(t))$ oscillates with frequency $\alpha=q+q^{-1}$.}
\label{fig:EUt}
\end{figure}

Using the explicit matrix elements of $U(t)$ and the permutation representation of $T_{13}$, one evaluates the trace directly (details in Appendix~\ref{app:OE_WZ}) and finds
\begin{equation}
\Tr\!\left(U^{\otimes 2}T_{13}U^{\dagger\otimes 2}T_{13}\right)
=
\frac{8}{(q^2+1)^4}\Big[(q^2+1)^4+\Delta(q,t)\Big],
\label{eq:trace_result_main}
\end{equation}
where
\begin{equation}
\Delta(q,t)=(q^2-1)^4 c^2+8q^2(q^2-1)^2 c+16q^4,\,\,\, c=\cos(\alpha t).
\end{equation}
Substituting into \eqref{eq:WZ_2qubit} yields
\begin{equation}
E\big(U(t)\big)
=
\frac12-\frac{1}{2}\frac{\Delta(q,t)}{(q^2+1)^4}.
\label{eq:E_closed_main}
\end{equation}
This passes the standard checks:
(i) $q=1$ gives $E(U(t))=0$ for all $t$;
(ii) $t=0$ gives $E(U(0))=0$;
(iii) for $q\neq1$, $E(U(t))$ is generically nonzero and oscillatory.

Equation \eqref{eq:E_closed_main} is our first structural result:
the non-cocommutative coproduct produces a strictly nontrivial operator entanglement in the two-qubit sector, even though the single-qubit Hamiltonian is undeformed.

\begin{figure}[h]
\centering
\includegraphics[width=1\linewidth]{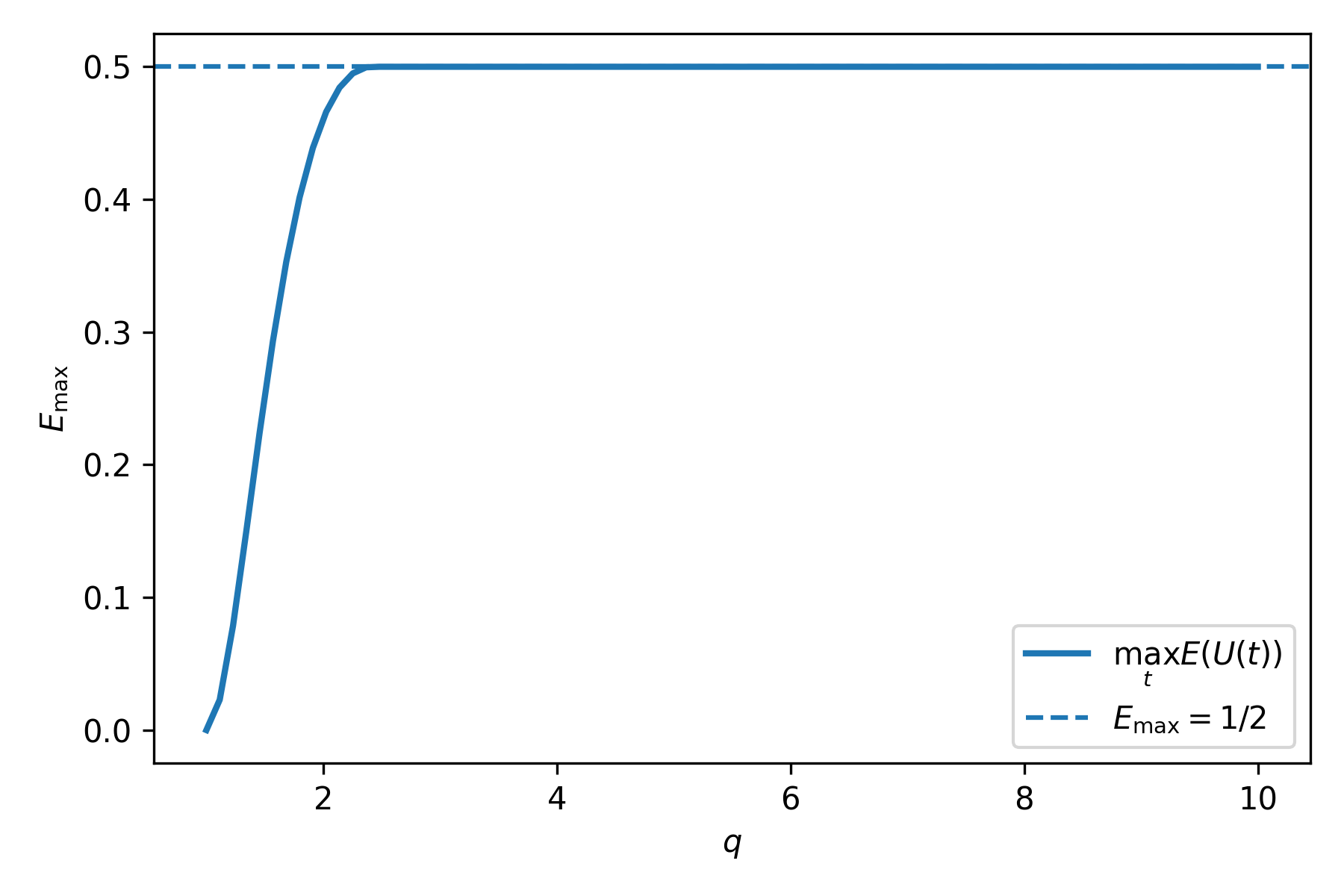}
\caption{Numerical maximization of $E(U(t))$ over $t$.
The asymptotic limit $E_{\max}\to \tfrac12$ as $q\to\infty$ is shown; the maximum is effectively saturated already for $q\gtrsim2$.}
\label{fig:Emax}
\end{figure}

A further structural consequence follows from the cubic identity \eqref{eq:cubic_main}.
For $\alpha\neq0$, the unital associative algebra generated by $H_{AB}$ is
\begin{equation}
\mathbb{C}[H_{AB}]=\mathrm{span}\{\mathds{1},H_{AB},H_{AB}^2\},
\end{equation}
so the unitary evolution remains confined to a three-dimensional operator manifold for all times.
The resulting entanglement dynamics is integrable in a strong algebraic sense: oscillations and revivals are exact, and there is no irreversible growth of correlations.
Numerically, the maximal operator entanglement approaches the two-qubit upper bound $E_{\max}=1/2$ already for $q\gtrsim2$ (Fig.~\ref{fig:Emax}).

This spin-$\tfrac12$ realization is a minimal instance of what we call \emph{noncommutative kinematical scrambling}:
non-cocommutativity of the coproduct enforces nontrivial mixing and operator entanglement, but the finite algebraic closure prevents genuine operator growth and excludes chaotic scrambling.
To sharpen this distinction we now compute the entangling power.

\section{Entangling power}
\label{sec:entpow}

Operator entanglement quantifies the intrinsic nonlocality of a unitary \emph{as an operator}, but it does not by itself determine how much entanglement is typically generated from \emph{product inputs}.
A complementary notion is the entangling power~\cite{PhysRevA.62.030301},
\begin{equation}
\mathrm{e}_p(U)=
\mathbb{E}_p\!\left[E\!\left(U\ket{\psi_A}\otimes\ket{\psi_B}\right)\right],
\label{eq:ep_def}
\end{equation}
where the average is over product states according to a distribution $p$ and $E(\cdot)$ is the linear entropy of the reduced density matrix.

For Haar-uniform averaging over product states ($p=p_0$), Wang and Zanardi obtained~\cite{PhysRevA.66.044303}
\begin{equation}
\mathrm{e}_{p_0}(U)=
\frac{d_Ad_B}{(d_A+1)(d_B+1)}
\Big[E(U)+E(US)-E(S)\Big],
\label{eq:ep_WZ}
\end{equation}
where $S$ is the physical swap on a single copy $AB$. $E(S)$ measures the static maximal nonlocality of the swap, and $E(US)$ measures how much scrambling is induced relative to swapping subsystems.

For two qubits ($d_A=d_B=2$),
\begin{equation}
\mathrm{e}_{p_0}(U)=\frac{4}{9}\left[E(U)+E(US)-\frac{3}{4}\right],
\label{eq:ep_2qb}
\end{equation}
with $E(S)=1-\frac{1}{d_Ad_B}=\frac{3}{4}$.

The quantity $E(US)$ can be expressed as~\cite{PhysRevA.66.044303}
\begin{equation}
E(US)=
1-\frac{1}{16}\,
\Tr\!\left(
U^{\otimes 2}\,T_{24}\,U^{\dagger\otimes 2}\,T_{13}
\right),
\label{eq:EUS_def}
\end{equation}
with $T_{13}$ and $T_{24}$ swapping $(1,3)$ and $(2,4)$, respectively.
For $U(t)=e^{-itH_{AB}(q)}$ a remarkable simplification occurs: inserting \eqref{eq:U_matrix} and using only $c^2+s^2=1$, all $q$- and $t$-dependence cancels (Appendix~\ref{app:tildeE_WZ}) and one finds
\begin{equation}
\Tr\!\left(
U^{\otimes 2}\,T_{24}\,U^{\dagger\otimes 2}\,T_{13}
\right)=4
\qquad \forall\,q>0,\ \forall\,t,
\end{equation}
hence
\begin{equation}
E\big(U(t)S\big)=1-\frac{4}{16}=\frac{3}{4}.
\label{eq:EUS_const}
\end{equation}
Substituting into \eqref{eq:ep_2qb} yields the compact relation
\begin{equation}
\mathrm{e}_{p_0}\!\left(U(t)\right)=\frac{4}{9}\,E\!\left(U(t)\right).
\label{eq:ep_slaved}
\end{equation}

Equation~\eqref{eq:ep_slaved} is our second structural result:
in this coproduct-generated two-qubit model the entangling power carries no independent dynamical content beyond the operator entanglement itself.
In particular, for $q=1$ one has $E(U(t))=\mathrm{e}_{p_0}(U(t))=0$ for all $t$, while for $q\neq1$ both are nonzero and oscillate with the same characteristic frequency $\alpha=q+q^{-1}$.

The oscillatory behaviour of $E(U(t))$ is a direct consequence of the finite-dimensional closure implied by \eqref{eq:cubic_main}.
Equivalently, the Heisenberg evolution preserves a low-dimensional invariant subspace in operator space: nested commutators do not proliferate, operator components undergo coherent rotations, and entanglement exhibits revivals rather than irreversible growth.
In this sense the present model realizes the typical behavior of Rabi oscillations \emph{in operator space}.

This behavior sharply contrasts with chaotic scrambling~\cite{Hosur:2015ylk,Chen:2018hjf, Garcia:2022abt}.
In generic nonintegrable dynamics the operator algebra generated by time evolution does not close on a small set; operators grow in support and complexity, the commutator tower expands rapidly, and entanglement measures typically show growth and saturation.
Here, the nonlocality induced by the noncocommutative coproduct produces nontrivial operator entanglement, but algebraic closure prevents genuine operator growth and excludes irreversible information spreading. We therefore refer to the qubit coproduct model as an instance of \emph{noncommutative kinematical scrambling}.

\section{Coproduct-induced, irreducible entanglement}

In the present construction $H_{AB}(q)$ is not an arbitrary two-qubit Hamiltonian: it is fixed by the coproduct and therefore by the symmetry-composition rule itself.

Local changes of basis on $\mathcal{H}_A\otimes\mathcal{H}_B$ correspond to unitary equivalences of the \emph{representation}.
However, the Hopf-algebraic rule for composing generators is part of the kinematics: it is encoded in the coproduct and is not altered by such local basis changes.
In particular, for $q\neq1$ the coproduct remains non-cocommutative.

For $q=1$, cocommutativity allows $H_{AB}(1)$ to be a sum of local terms and $U(t)$ to factorize, giving $E(U(t))=0$.
For $q\neq1$, the deformation survives at the two-qubit level precisely through the nonlocal couplings dictated by $\Delta$, leading to $E(U(t))\neq0$ at generic times.
Therefore the nonzero operator entanglement of $U(t)$ for $q\neq1$ is not an artefact of a particular gate decomposition: it is a structural nonlocality imposed by the quantum-group symmetry.
Any faithful realization of $U_q(\mathfrak{su}(2))$ on two sites inherits this coproduct-induced operator entanglement in its composite generators.

\section{Conclusions and outlook}

We have demonstrated, in a fully explicit and exactly solvable two-qubit model, that Hopf-algebra deformations of symmetry can enforce an intrinsic layer of operator entanglement at the level of subsystem composition.
In the $U_q(\mathfrak{su}(2))$ example the deformation is invisible on a single qubit but resurfaces through the coproduct in the two-qubit sector, producing nonzero, oscillatory operator entanglement for any $q\neq1$.

Two structural features underlie the simplicity of the model.
First, the coproduct fixes the composite generator and hence the nonlocal couplings in $H_{AB}(q)$, making the entanglement content a kinematical imprint of non-cocommutativity.
Second, in the fundamental representation $H_{AB}(q)$ obeys the cubic identity $H^3=\alpha^2H$ with $\alpha=q+q^{-1}$, implying closure of $\mathbb{C}[H_{AB}]$ and exact integrability.
As a consequence, operator entanglement exhibits revivals and cannot develop into chaotic scrambling; correspondingly, for Haar-uniform product inputs the entangling power is entirely determined by $E(U(t))$ via Eq.~\eqref{eq:ep_slaved}.

The present analysis should thus be viewed as a minimal integrable instance in which non-cocommutativity seeds a baseline of operator entanglement.
A natural next step is to move beyond the accidental qubit: for higher representations $j\ge1$ the single-site Hamiltonian $H(q)$ acquires genuine $q$-dependence, low-degree polynomial closure is not expected, and the composite dynamics can explore much larger operator algebras.
In many-body settings built from iterated coproducts, the same Hopf-algebraic mechanism identified here may combine with nonintegrable dynamics to yield substantial operator growth and scrambling~\cite{Andreadakis:2024nal,PhysRevLett.126.030601}. In more general quantum-group settings (higher spins, higher rank, or $\kappa$-deformed symmetries) the coproduct-enforced baseline of operator entanglement identified here may interact with nonintegrable dynamics to yield richer operator growth, potentially affecting circuit complexity and the simulation of quantum-gravitational effective dynamics~\cite{Ballesteros:2025dbv}.

More broadly, our results suggest a shift of viewpoint: in the presence of non-commutative deformations of symmetries, entanglement need not be purely dynamical, but may be partly encoded in the algebraic rules by which subsystems are composed.
This provides a concrete bridge between quantum-group symmetries and operational notions of nonlocality relevant to quantum-information processing in quantum-spacetime scenarios.

Finally, we shall remark that while we have focused on operator entanglement and entangling power as probes of coproduct-induced scrambling, an important and conceptually complementary direction concerns the role of non-stabilizerness (magic) as a resource for quantum dynamics~\cite{Emerson:2013zse, PhysRevLett.128.050402}. In this respect, the present coproduct-induced dynamics provides a natural arena to investigate how Hopf-algebraic structures and quantum symmetries constrain or generate magic, extending recent analyses of symmetry-protected magic and symmetry-resolved non-stabilizerness~\cite{Cepollaro:2024qln,cepollaro2025stabilizerentropysubspaces}.

\section*{Acknowledgments}
M.A. acknowledges support from the INFN Iniziativa Specifica QUAGRAP and from the European COST Action BridgeQG CA23130.
G.C. acknowledges support from the INFN Iniziativa Specifica GeoSymQFT. Both authors acknowledge support from the European COST Action CaLISTA CA21109.

%

\onecolumngrid

\appendix

\section{Single-spin representation and $q$-exponentials}
\label{app:single_spin}

We work in the fundamental (spin-$\tfrac12$) representation with basis
\begin{equation}
\ket{0}=
\begin{pmatrix}1\\0\end{pmatrix},
\qquad
\ket{1}=
\begin{pmatrix}0\\1\end{pmatrix}.
\end{equation}
Then
\begin{equation}
J_z=\frac12
\begin{pmatrix}
1&0\\
0&-1
\end{pmatrix},
\qquad
J_+=
\begin{pmatrix}
0&1\\
0&0
\end{pmatrix},
\qquad
J_-=
\begin{pmatrix}
0&0\\
1&0
\end{pmatrix},
\qquad
\sigma_x=J_+ + J_-.
\end{equation}
The $q$-exponentials are
\begin{equation}
q^{J_z}=
\begin{pmatrix}
q^{1/2}&0\\
0&q^{-1/2}
\end{pmatrix},
\qquad
q^{J_z/2}=
\begin{pmatrix}
q^{1/4}&0\\
0&q^{-1/4}
\end{pmatrix}.
\end{equation}
The single-spin Hamiltonian is
\begin{equation}
H(q)=q^{J_z/2}(J_+ + J_-)q^{J_z/2},
\end{equation}
and one checks immediately
\begin{equation}
H(q)=
\begin{pmatrix}
q^{1/4}&0\\
0&q^{-1/4}
\end{pmatrix}
\begin{pmatrix}
0&1\\
1&0
\end{pmatrix}
\begin{pmatrix}
q^{1/4}&0\\
0&q^{-1/4}
\end{pmatrix}
=
\begin{pmatrix}
0&1\\
1&0
\end{pmatrix}
=\sigma_x,
\qquad \forall\,q.
\end{equation}

\section{Coproduct-derived two-qubit Hamiltonian}
\label{app:hab}

We derive the two-qubit Hamiltonian from the coproduct and clarify its relation to the compact form \eqref{eq:HAB_def}.
Using the coproduct \eqref{eq:coprJpm} and the homomorphism property $\Delta(AB)=\Delta(A)\Delta(B)$, one has
\begin{equation}
\Delta\!\left(q^{J_z/2}\right)=q^{\Delta(J_z)/2}=q^{J_z/2}\otimes q^{J_z/2}.
\label{eq:coproduct_qJz2}
\end{equation}
The group-like property $\Delta(K)=K\otimes K$ for $K=q^{J_z}$ is standard in the Chevalley presentation of $U_q(\mathfrak{su}(2))$ (see e.g.\ \cite{Chari:1994pz,Majid:1996kd,Klimyk:1997eb}), and implies \eqref{eq:coproduct_qJz2} by functional calculus and the homomorphism property of $\Delta$.

Therefore,
\begin{equation}
\Delta(H)=\Delta\!\left(q^{J_z/2}\right)\Delta(J_+ + J_-)\Delta\!\left(q^{J_z/2}\right).
\end{equation}
Evaluating in the basis $\{\ket{00},\ket{01},\ket{10},\ket{11}\}$ yields a matrix unitarily equivalent to the form used in the main text; after a basis permutation (swapping $\ket{01}\leftrightarrow\ket{10}$) one obtains precisely
\begin{equation}
H_{AB}(q)=\sigma_x\otimes\mathds{1}+q^{2J_z}\otimes\sigma_x,
\end{equation}
i.e.\ Eq.~\eqref{eq:HAB_def}.

\section{Exponential map $U(t)=e^{-itH_{AB}(q)}$}
\label{app:uni}

With $H\equiv H_{AB}(q)$, the cubic identity $H^3=\alpha^2H$ with $\alpha=(q^2+1)/q$ implies
\begin{equation}
e^{-itH}
=\mathds{1}-i\frac{\sin(\alpha t)}{\alpha}H+\frac{\cos(\alpha t)-1}{\alpha^2}H^2,
\end{equation}
which leads directly to the explicit matrix \eqref{eq:U_matrix} given in the main text.

\section{Operator entanglement from the coproduct unitary}
\label{app:OE_WZ}

We compute the trace in \eqref{eq:WZ_2qubit} for $U(t)$ in \eqref{eq:U_matrix}.
Defining
\begin{equation}
\mathcal{S}(U)\coloneqq
\Tr\!\left(U^{\otimes 2}T_{13}U^{\dagger\otimes 2}T_{13}\right),
\end{equation}
a direct evaluation gives
\begin{equation}
\mathcal{S}\big(U(t)\big)
=
\frac{8}{(q^2+1)^4}\Big[
(q^2-1)^4 c^2 + 8q^2(q^2-1)^2 c + (q^2+1)^4 + 16q^4
\Big],
\end{equation}
with $c=\cos(\alpha t)$, $\alpha=(q^2+1)/q$.
Substituting into \eqref{eq:WZ_2qubit} yields the closed form \eqref{eq:E_closed_main}.

\section{The mixed invariant $\widetilde E(U)$ and entangling power}
\label{app:tildeE_WZ}

Wang and Zanardi define
\begin{equation}
\widetilde E(U)=
1-\frac{1}{16}\Tr\!\left(U^{\otimes 2}T_{24}U^{\dagger\otimes 2}T_{13}\right)
\end{equation}
for two qubits.
For $U(t)$ in \eqref{eq:U_matrix}$,$ one finds
\begin{equation}
\Tr\!\left(U^{\otimes 2}T_{24}U^{\dagger\otimes 2}T_{13}\right)=4
\qquad \forall\,q>0,\ \forall\,t,
\end{equation}
hence $\widetilde E(U(t))=3/4$ and therefore Eq.~\eqref{eq:ep_slaved}.

\end{document}